\documentclass[aps,pra,twocolumn,floatfix,longbibliography,showpacs]{revtex4-1} %nofootinbib is not enabled

\usepackage{amsmath,amsfonts,amssymb,bm}
\usepackage{dcolumn}
\usepackage{braket}
\usepackage[final]{graphicx}
\usepackage{tabularx}
\usepackage{graphicx}

\usepackage{dsfont}
\def\dblone{\mathds{1}}

\begin{document}

\title{A Rotating-Frame Perspective on High-Harmonic Generation of Circularly Polarized Light}

\author{Daniel M. Reich}
\affiliation{Department of Physics and Astronomy, Aarhus University, DK-8000 Aarhus C, Denmark}
\author{Lars Bojer Madsen}
\affiliation{Department of Physics and Astronomy, Aarhus University, DK-8000 Aarhus C, Denmark}

\pacs{32.80.Wr, 42.25.Ja, 42.50.Tx, 42.65.Ky}

\begin{abstract}
We employ a rotating frame of reference to elucidate high-harmonic generation of circularly polarized
light by bicircular driving fields. In particular, we show how the experimentally observed
circular components of the high-harmonic spectrum can be directly related to the corresponding
quantities in the rotating frame. Supported by numerical simulations of the time-dependent Schr\"{o}dinger equation, we
deduce an optimal strategy for maximizing the cutoff in the high-harmonic plateau while
keeping the two circular components of the emitted light spectrally distinct. Moreover, we show
how the rotating-frame picture can be more generally employed for elliptical drivers.
Finally, we point out how circular and elliptical driving fields show a near-duality to static electric and
static magnetic fields in a rotating-frame description. This demonstrates how
high-harmonic generation of circularly polarized light under static electromagnetic fields can be
emulated in practice even at static field strengths beyond current experimental capabilities.
\end{abstract}

\maketitle

\section{Introduction}

High-harmonic generation (HHG) of circularly polarized light was studied
theoretically already in the 1990s~\cite{Eichmann1995,Long1995,Becker1999,Milosevic2000,Milosevic2000a}
but experimental demonstrations were not successful until very recently~\cite{Fleischer2014,Kfir2015}.
The experimentally obtained high-harmonic spectra for such fields,
including extensions towards elliptical drivers~\cite{Fleischer2014}, inspired
theoretical efforts to provide simple explanations for the observed
features, in particular with respect to conservation of photon angular
momenta and corresponding selection rules~\cite{Pisanty2014,Milosevic2015}.
Generating isolated circularly polarized high harmonics has been the
subject of further theoretical~\cite{Medisauskas2015} and experimental~\cite{Hickstein2015}
studies. Furthermore, it was shown recently that high-harmonic spectra
with circular polarization can be made to reach even into the soft
X-ray regime~\cite{Fan17112015}.

At the base of most schemes for generation of high-harmonic circular
light lies the use of a driving field that is a superposition of two
copropagating but counter-rotating circular drivers with different frequencies.
It has recently also been shown that counter-rotating drivers with the
same frequency can be used when employing a non-collinear driving scheme~\cite{Hickstein2015}.
In a somewhat different context, namely in the presence of additional
static electromagnetic fields, HHG with such counter-rotating fields was also studied in the
2000s~\cite{Milosevic1999,Bandrauk2003}. However, these studies
neither considered the generated high-harmonic light in terms of its
circularly polarized components nor did they report significant effects
on the spectrum for readily available static magnetic and electric
field strengths.

In this paper we show how a great deal of understanding of HHG
with non-linearly polarized driving fields can be obtained
by going to a rotating frame. Focusing on the paradigmatic case of
counter-rotating bicircular drivers we illustrate how this allows to obtain a
surprisingly simple characterization that even enables the formulation
of optimal strategies for obtaining high-order harmonics with well-characterized
circular polarization.

We begin by describing in Sec.~\ref{sec:linearising rotating frame}
how a rotating frame of reference can be employed to obtain a single
linearly polarized driving field from a bicircular driver in the lab
frame. In Sec.~\ref{sec:hhg_rot_transform} we show that there is
a direct connection between the high-harmonic spectra in rotating
frames and the lab frame when considering the circular components
of the emitted light. This allows us to illustrate in Sec.~\ref{sub:average frequency apply}
that the most striking properties of the generation of circularly
polarized high-harmonics can be very easily explained utilizing the
linearizing rotating frame introduced in Sec.~\ref{sec:linearising rotating frame}.
To emphasize this point even further we show in Sec.~\ref{sub:bicirc optimise}
how the rotating-frame description allows to formulate an optimal
experimental strategy towards obtaining a high-harmonic plateau in
the emission spectrum that is as far extended as possible while still
separating the left and right circularly-polarized components energetically.
We support these findings with numerical simulations of HHG
on the single-atom level by solving the time-dependent Schr\"{o}dinger equation for a two-dimensional
model atom using the single-active-electron approximation.
We finish our discussion by presenting two particular perspectives on the
rotating-frame description: 
In Sec.~\ref{sec:perspectives I} we consider a more general setup involving 
elliptical driving fields and in Sec.~\ref{sec:perspectives II}
we show how the study of HHG in rotating frames
is strongly connected to questions concerning HHG in the presence
of static electromagnetic fields. Finally, Sec.~\ref{sec:conclusions} concludes.

Atomic units are used throughout unless otherwise noted.

\section{The Linearizing Rotating Frame for Bicircular Drivers\label{sec:linearising rotating frame}}

We consider the Hamiltonian of an axially symmetric field-free Hamiltonian
$H_{0}$ under the influence of the electric field of two counter-rotating
circularly polarized laser pulses with envelope $F_{0}\left(t\right)$
and frequencies $\omega_{1},\omega_{2}$,
\begin{eqnarray}
  \nonumber H\left(t\right)&=&H_{0}+F_{0}\left(t\right)\left[x\cos\left(\omega_{1}t\right)+y\sin\left(\omega_{1}t\right)\right. \\
                               & &\left.+x\cos\left(\omega_{2}t\right)-y\sin\left(\omega_{2}t\right)\right]\,.\label{eq:bicirc_start}
\end{eqnarray}
The Hamiltonian (\ref{eq:bicirc_start}) models an atom or a linear
molecule aligned along the $z$-axis.

We will perform a unitary transformation on this Hamiltonian, first
applied to HHG processes in Ref.~\cite{Bandrauk2003}
and first mentioned in the explicit context of generation of circularly polarized
high harmonics in Ref.~\cite{Yuan2015}, given by the following expression,
\begin{equation}
U\left(t\right)=e^{-i\alpha tL_{z}}\,,\label{eq:unitary}
\end{equation}
with $L_{z}$ being the operator of angular momentum corresponding
to rotation around the $z$-axis. Hence, Eq.~(\ref{eq:unitary})
represents a rotation in the $x,y$-plane with angular frequency $-\alpha$.
The rotating Hamiltonian reads
\begin{eqnarray*}
H'\left(t\right) & = & U\left(t\right)HU^{\dagger}\left(t\right)-iU\left(t\right)\frac{\partial}{\partial t}U^{\dagger}\left(t\right)\,.
\end{eqnarray*}
Clearly $U\left(t\right)H_{0}U^{\dagger}\left(t\right)=H_{0}$ due to the axial symmetry assumed
for $H_{0}$. Furthermore
\begin{eqnarray*}
U\left(t\right)xU^{\dagger}\left(t\right) & = & x\cos\left(\alpha t\right)+y\sin\left(\alpha t\right)\,,\\
U\left(t\right)yU^{\dagger}\left(t\right) & = & y\cos\left(\alpha t\right)-x\sin\left(\alpha t\right)\,,
\end{eqnarray*}
which allows us to transform the interaction term $H_{\text{int}}\left(t\right)=H\left(t\right)-H_{0}$ in Eq.~(\ref{eq:bicirc_start})
as follows,
\begin{widetext}
\begin{eqnarray*}
U\left(t\right)H_{\text{int}}\left(t\right)U^{\dagger}\left(t\right) & = & F_{0}\left(t\right)\left[x\cos\left(\omega_{1}t\right)\cos\left(\alpha t\right)+y\cos\left(\omega_{1}t\right)\sin\left(\alpha t\right)+y\sin\left(\omega_{1}t\right)\cos\left(\alpha t\right)-x\sin\left(\omega_{1}t\right)\sin\left(\alpha t\right)\right.\\
 &  & \left.+x\cos\left(\omega_{2}t\right)\cos\left(\alpha t\right)+y\cos\left(\omega_{2}t\right)\sin\left(\alpha t\right)-y\sin\left(\omega_{2}t\right)\cos\left(\alpha t\right)+x\sin\left(\omega_{2}t\right)\sin\left(\alpha t\right)\right]\,.
\end{eqnarray*}
Using the addition theorems of sine and cosine one arrives at the expression
\begin{equation}
U\left(t\right)H_{\text{int}}\left(t\right)U^{\dagger}\left(t\right)=F_{0}\left(t\right)\left(x\cos\left(\left[\omega_{1}+\alpha\right]t\right)+y\sin\left(\left[\omega_{1}+\alpha\right]t\right)+x\cos\left(\left[\omega_{2}-\alpha\right]t\right)-y\sin\left(\left[\omega_{2}-\alpha\right]t\right)\right)\,.\label{eq:general_rotation_ham}
\end{equation}
\end{widetext}
By choosing
\begin{equation}
\alpha=\frac{\omega_{2}-\omega_{1}}{2}\label{eq:linearising alpha}
\end{equation}
and defining the average frequency 
\begin{equation}
\tilde{\omega}\equiv\frac{\omega_{1}+\omega_{2}}{2}=\omega_{1}+\alpha=\omega_{2}-\alpha\,.\label{eq:average frequency}
\end{equation}
we observe that the sine terms in Eq.~(\ref{eq:general_rotation_ham}) cancel and only a linearly polarized
field with frequency $\tilde{\omega}$ remains. Hence we obtain for
the total Hamiltonian in the rotating frame,
\begin{equation}
H'\left(t\right)=H_{0}+\alpha L_{z}+2F_{0}\left(t\right)x\cos\left(\tilde{\omega}t\right)\,.\label{eq:average_frequency_rotated_ham}
\end{equation}
We will mostly focus on the case $\omega_{2} > \omega_{1}$ whence by
Eq.~(\ref{eq:linearising alpha}) $\alpha > 0$ follows.

We conclude that the dynamics of two counter-rotating circularly polarized
driving fields can be interpreted as a single linearly polarized driver
with double the field strength at the mean frequency. In the rotating frame
an additional angular momentum term appears, which we will call coriolis term in accordance
with Ref.~\cite{Yuan2015}, proportional to half the difference frequency.
For two co-rotating circularly polarized drivers the result is almost
identical and can be obtained by substituting $\omega_{2}\rightarrow-\omega_{2}$.
The linearly polarized field in this case has a frequency
equal to half the difference in frequency between the circular drivers
while the coriolis term will now be equal to $\alpha' L_{z}$ with
$\alpha'=\frac{\omega_{1}+\omega_{2}}{2}$ corresponding to the mean
frequency.

\section{Translating Harmonic Spectra between Lab and Rotating Frames\label{sec:hhg_rot_transform}}

How does one translate the harmonic signal of an atom or molecule
from the rotating frame back to the lab frame? The high-harmonic signal
in the lab frame has regularly been associated with the square of
the Fourier transform of the dipole expectation value~\cite{Corkum1993,Bandrauk2008,Zhao2008},
e.g., for the harmonic signal in $x$-direction one obtains
\begin{equation*}
S_{x}^{\text{lab}}\left(\omega\right)=\left|\int D_{x}^{\text{lab}}\left(t\right)e^{-i\omega t}\ dt\right|^{2}=\left|\int\left\langle x\right\rangle _{\text{lab}}e^{-i\omega t}\ dt\right|^{2}\,,
\end{equation*}
where the expectation value $\left\langle \cdot\right\rangle _{\text{lab}}$ is taken with respect to the time-dependent wave function
in the lab frame, $\ket{\psi\left(t\right)}$. 
Note that we already absorbed the minus sign from the dipole expectation value
due to the electronic charge into the modulus.
The wave function in
the rotating frame is given by $U\left(t\right)\Ket{\psi\left(t\right)}$
with $U\left(t\right)$ given by Eq.~(\ref{eq:unitary}). Using $U\left(t\right)U^{\dagger}\left(t\right)=U^{\dagger}\left(t\right)U\left(t\right)=\dblone$
we obtain the expression
\begin{eqnarray*}
  D_{x}^{\text{lab}}\left(t\right) & = & \left\langle x\right\rangle _{\text{lab}}=\left\langle \psi\left(t\right)|x|\psi\left(t\right)\right\rangle \\
                                   & = & \left\langle \psi\left(t\right)|U^{\dagger}\left(t\right)U\left(t\right)xU^{\dagger}\left(t\right)U\left(t\right)|\psi\left(t\right)\right\rangle \\
                                   & = & \left\langle x\right\rangle _{\text{rot}}\cos\left(\alpha t\right)+\left\langle y\right\rangle _{\text{rot}}\sin\left(\alpha t\right)\,,
\end{eqnarray*}
where $\left\langle \cdot\right\rangle _{\text{rot}}$ denotes expectation
values with respect to the time-dependent wave function in the rotating
frame. The analogous expressions for the $y$-direction, $D_{y}^{\text{lab}}\left(t\right)$,
as well as the right-circular component, $D_{+}^{\text{lab}}\left(t\right)$,
and left-circular component, $D_{-}^{\text{lab}}\left(t\right)$,
are given by
\begin{eqnarray*}
  D_{y}^{\text{lab}}\left(t\right) & = & -\left\langle x\right\rangle _{\text{rot}}\sin\left(\alpha t\right)+\left\langle y\right\rangle _{\text{rot}}\cos\left(\alpha t\right)\,,\\
  D_{+}^{\text{lab}}\left(t\right) & = & \frac{1}{\sqrt{2}}\left(D_{x}^{\text{lab}}\left(t\right)+iD_{y}^{\text{lab}}\left(t\right)\right)\\
                                   & = & \frac{1}{\sqrt{2}}\left(\left\langle x\right\rangle _{\text{rot}}\cos\left(\alpha t\right)+\left\langle y\right\rangle _{\text{rot}}\sin\left(\alpha t\right)\right. \\
                                   &   & \left.+i\left\langle y\right\rangle _{\text{rot}}\cos\left(\alpha t\right)-i\left\langle x\right\rangle _{\text{rot}}\sin\left(\alpha t\right)\right)\,,\\
  D_{-}^{\text{lab}}\left(t\right) & = & \frac{1}{\sqrt{2}}\left(D_{x}^{\text{lab}}\left(t\right)-iD_{y}^{\text{lab}}\left(t\right)\right)\\
                                   & = & \frac{1}{\sqrt{2}}\left(\left\langle x\right\rangle _{\text{rot}}\cos\left(\alpha t\right)+\left\langle y\right\rangle _{\text{rot}}\sin\left(\alpha t\right)\right. \\
                                   &   & \left.-i\left\langle y\right\rangle _{\text{rot}}\cos\left(\alpha t\right)+i\left\langle x\right\rangle _{\text{rot}}\sin\left(\alpha t\right)\right)\,.
\end{eqnarray*}
For the circular components the expressions can be significantly simplified
by using cylindrical coordinates which leads to the expressions
\begin{eqnarray*}
D_{+}^{\text{lab}}\left(t\right) & = & \frac{1}{\sqrt{2}}\left\langle \rho e^{i\left(\varphi-\alpha t\right)}\right\rangle _{\text{rot}}\,,\\
D_{-}^{\text{lab}}\left(t\right) & = & \frac{1}{\sqrt{2}}\left\langle \rho e^{i\left(\varphi+\alpha t\right)}\right\rangle _{\text{rot}}\,,
\end{eqnarray*}
where $x=\rho\cos\varphi$ and $y=\rho\sin\varphi$. This equation
alludes to a direct connection between the harmonic spectra obtained
in the rotating frame and the high-harmonic spectra in the lab frame.
In fact, a short calculation shows that
\begin{eqnarray}
S_{+}^{\text{lab}}\left(\omega\right) & = & \left|\int D_{+}^{\text{lab}}\left(t\right)e^{-i\omega t}\ dt\right|^{2}\nonumber \\
 & = & \left|\int\left\langle \frac{1}{\sqrt{2}}\rho e^{i\varphi}\right\rangle _{\text{rot}}e^{-i\left(\omega+\alpha t\right)}\ dt\right|^{2}\nonumber \\
 & = & S_{+}^{\text{rot}}\left(\omega+\alpha\right)\nonumber\,,\\
S_{+}^{\text{lab}}\left(\omega-\alpha\right) & = & S_{+}^{\text{rot}}\left(\omega\right)\label{eq:plus signal}
\end{eqnarray}
and analogously
\begin{eqnarray}
S_{-}^{\text{lab}}\left(\omega\right) & = & S_{-}^{\text{rot}}\left(\omega-\alpha\right)\nonumber\,,\\
S_{-}^{\text{lab}}\left(\omega+\alpha\right) & = & S_{-}^{\text{rot}}\left(\omega\right)\,.\label{eq:minus signal}
\end{eqnarray}
We conclude that the right (left) circular signal in the lab frame
is obtained via the right (left) circular signal in the rotating frame
shifted in frequency by $\alpha$ to the left (to the right). We 
emphasize that this statement is general and no assumptions on
$\alpha,\ \omega$ or the particular structure of $H_{0}$ (with the
exception of its axial symmetry) have been made. This simple relationship
between the circular spectral components in lab frame and rotating
frame allows to directly apply any insights obtained about the high
harmonic spectra in the rotating frame when discussing the experimentally
observed spectra in the lab frame.

Finally, it should be pointed out that often the high-harmonic signal
is associated with the Fourier transform of the dipole acceleration
rather than that of the dipole moment \cite{Ivanov1996,Lein2002,Kamta2005}.
In Appendix \ref{sec:Dipole acceleration} we show that computing
the corresponding expectation values via Ehrenfest's theorem with
a slight modification allows to preserve the validity of Eqs.~(\ref{eq:plus signal})
and (\ref{eq:minus signal}). The result from a formulation in terms of the dipole velocity
can be derived by using the expressions from the dipole
or the dipole acceleration with an appropriate scaling of
the signal in terms of frequency \cite{Baggesen2011}.
Throughout the remainder of this work we compute high-harmonic
spectra via the dipole acceleration using Ehrenfest's theorem.

\section{Understanding HHG Spectra from Analyses in Rotating Frames\label{sub:average frequency apply}}

In the linearizing rotating frame, the high-harmonic signal of right-circular,
respectively left-circular, polarization is very similar
to that of the system under a linearly polarized driver at the mean
frequency, shifted with half the difference frequency up, respectively
down, cf.~Eqs.~(\ref{eq:linearising alpha}), (\ref{eq:average frequency}), (\ref{eq:plus signal}) and (\ref{eq:minus signal}).
The only difference to the well-studied HHG under a
linearly polarized driver lies in the additional coriolis term $\alpha L_{z}$
in Eq.~(\ref{eq:average_frequency_rotated_ham}) that influences
the system dynamics.

\begin{figure}[pt]
\centerline{\includegraphics[width=\columnwidth]{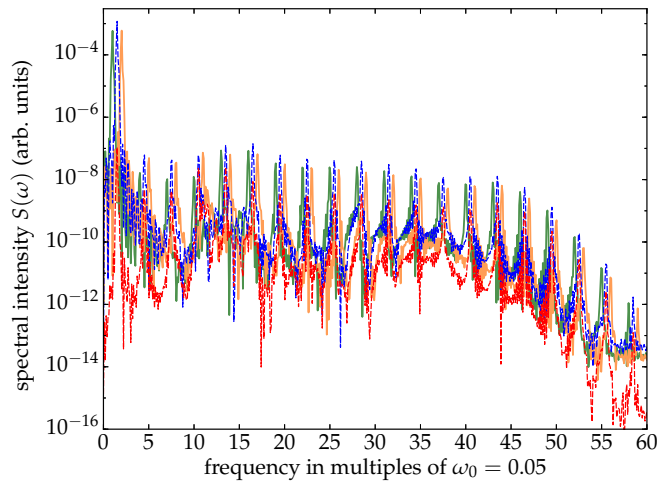}}
\caption{(color online) High-harmonic spectrum for two counter-rotating
circular drivers with $\omega_{1}=\omega_{0}=0.05$,
$\omega_{2}=2\omega_{1}=0.1$ and amplitude $F_{0}=0.05$ in the lab frame.
The drivers are trapezoidally shaped with ramp time of 2 cycles, $T_{\text{ramp}}=\frac{4\pi}{\omega_{0}}$
and plateau time of 5 cycles, $T_{\text{plateau}}=\frac{10\pi}{\omega_{0}}$.
The solid green (dark-shaded) curve is the right-circularly polarized signal in the lab
frame, the solid orange curve (light-shaded) is the left-circularly polarized signal in
the lab frame. The dashed blue (dark-shaded) curve is the signal in $x$-direction
in the rotating frame (with $\alpha=0.025$) while the
dashed red (light-shaded) curve is the signal in $y$-direction in the rotating frame. See
Fig.~\ref{fig:peakana_zoom} for a zoom-in.}
\label{fig:peakana}
\end{figure}

To illustrate how a great deal of insight regarding HHG of circularly polarized light can be obtained by thinking
in terms of rotating frames, we perform numerical simulations
of the TDSE for a two-dimensional model atom in both lab frame and rotating frame.
We will continue to call the prefactor of the coriolis term, corresponding to the 
negative rotation frequency of the rotating frame, $\alpha$. We look
at the most prevalent example which has been studied by recent experiments
- a counter-rotating circular driving scheme using a first and second
harmonic~\cite{Fleischer2014,Kfir2015}, i.e., $\omega_{1}=\omega_{0}$ and $\omega_{2}=2\omega_{0}$,
cf.~Eq.~(\ref{eq:bicirc_start}).
Transforming to a rotating frame
with $\alpha=0.5\omega_{0}$ [Eq.~(\ref{eq:linearising alpha})]
we obtain $\tilde{\omega}=1.5\omega_{0}$
[Eq.~(\ref{eq:average frequency})]. 
Because our arguments do not depend on a particular atomic or molecular species we were able to use
identical simulation parameters as in a recent theoretical study
that involved numerical simulations with a TDSE in two dimensions for such a set of bicircular counter-rotating
drivers, cf.~Table I in Ref.~\cite{Medisauskas2015}. 
The only numerical difference in our simulation consists in a slightly different complex
absorber and us employing a Chebyshev propagator \cite{Tal‐Ezer1984} with a Fourier method
for the kinetic energy.
Matching the parameters to the $s$ ground state calculation performed in \cite{Medisauskas2015} we
employ the screened potential $V\left(x,y\right)=V\left(\rho\right)=\frac{1}{\sqrt{\rho^{2}+0.1195}}$
leading to an ionization potential of $I_{p}=0.7935$. 
This allowed us to confirm the adequacy of our numerics by using our code
to reproduce Fig.~1(a) of Ref.~\cite{Medisauskas2015} in very good agreement
\footnote{Please refer to the most recent version of this figure which can be
downloaded from the URL \mbox{http://staff.mbi-berlin.de/medisaus}.}.

With this set of parameters we now turn to analyze a high-harmonic
spectrum in a rotating frame with $\alpha=0.025$. The driving field in this frame is
given by a trapezoidally shaped laser
pulse with frequency $\tilde{\omega}=0.075$ corresponding to a wavelength
of $\sim911\ \text{nm}$. The laser pulse is linearly polarized
along the $x$-direction and its maximal electric field strength is
given by $2F_{0}=0.1$. The linear ramp-up and ramp-down time
is given by $T_{\text{ramp}}=251.33$ and the plateau width is given
by $T_{\text{ramp}}=628.32$. 
In the lab frame this corresponds to a right-circularly
polarized driver superimposed with a counter-rotating left-circularly polarized
driver with frequencies $\tilde{\omega}-\alpha=0.05=\omega_{1}=\omega_{0}$, respectively
$\tilde{\omega}+\alpha=0.1=\omega_{2}=2\omega_{0}$, both with peak amplitude $F_{0}=0.05$. 

\begin{figure}[pt]
\centerline{\includegraphics[width=\columnwidth]{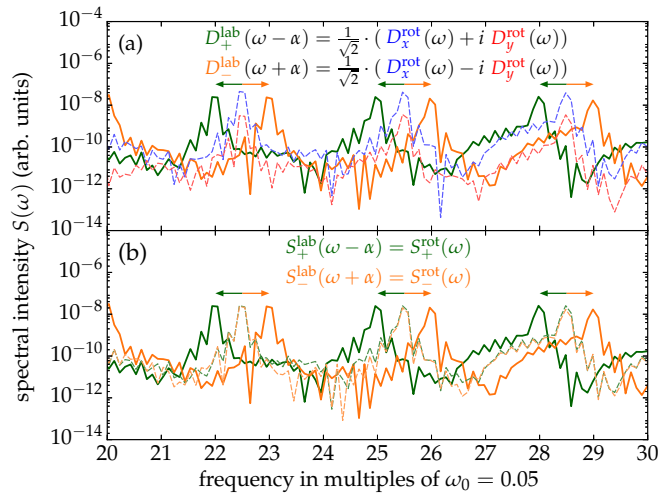}}
\caption{(color online) (a) Zoomed-in high-harmonic spectra from Fig.~\ref{fig:peakana}
illustrating the spectral shifts between rotating frame and lab frame.
(b) Same as (a) but with rotating frame spectra also shown in terms
of right- and left-circular components. The equations above the curves show the
relations discussed in Sec.~\ref{sec:hhg_rot_transform} between 
(a) dipole amplitudes and (b) spectral intensities in lab and rotating frame.}
\label{fig:peakana_zoom}
\end{figure}

In the rotating frame we can formulate the well-known selection rules of HHG driven by
a linearly polarized pulse, i.e., we obtain a signal only at odd multiples
of $\tilde{\omega}$. Note that these selection rules are not altered
by the coriolis term since it neither breaks the inversion symmetry
nor the conservation of $L^{2}$ and $L_{z}$. In the lab frame we
obtain accordingly by Eqs.~(\ref{eq:plus signal}) and (\ref{eq:minus signal})
a signal from a right, respectively left, circularly
polarized field at frequencies $\left[\left(2n+1\right)\tilde{\omega}-\alpha\right]$
and $\left[\left(2n+1\right)\tilde{\omega}+\alpha\right]$ ($n\in\mathbb{N}_0$)
corresponding to $\left(3n+1\right)\omega_{0}$ and $\left(3n-1\right)\omega_{0}$
which is consistent with the selection rules obtained in experiments
\cite{Fleischer2014,Hickstein2015} and previous theoretical studies \cite{Alon1998,Ceccherini2001,Nilsen2002}.
This behavior is confirmed by our numerical results, cf.~Figs.~\ref{fig:peakana}
and \ref{fig:peakana_zoom}. If we ignore the coriolis term in the
rotating frame we have emission purely in the $x$-direction and as such
the height of the two opposite circularly polarized peaks originating from
a given (linearly polarized) harmonic in the rotating frame would
be equal. However, the coriolis term does have a non-vanishing effect
on the direction of the emission in the rotating frame and there is
a visible contribution in the $y$-direction which can be directly associated
with the difference in peak heights. For most peaks the contribution
from the $y$-direction in the rotating frame for this particular set
of parameters is about one order of magnitude smaller than the contribution
from the $x$-direction. This explains the mostly close peak heights of neighboring
left- and right-circular harmonics in the lab frame, cf.~Figs.~\ref{fig:peakana}
and \ref{fig:peakana_zoom}.

We can even go a step further and understand the temporal generation
process of the circularly polarized light from the perspective of
the rotating frame. The highest ionization probability, which according
to the three-step model \cite{Corkum1993,Schafer1993,Lewenstein1994} forms the starting 
point of HHG, occurs around the peaks of the electric field. In the
rotating frame this means we have bursts of linearly polarized light
at minima and maxima of the electric field with frequency $\tilde{\omega}=1.5\omega_{0}$.
In terms of the period of the first harmonic driver $T=\frac{2\pi}{\omega_{0}}$
we consequently obtain bursts of linearly polarized high-harmonic
radiation at $t=0,\frac{2\pi}{3\omega_{0}},\frac{4\pi}{3\omega_{0}},\frac{2\pi}{\omega_{0}},\ldots=0,\frac{T}{3},\frac{2T}{3},T,\ldots$.
As discussed above, at least for moderately small coriolis terms,
we can approximate all emission in the rotating frame to take place
along the $x$-direction. At $t=0$ both frames are identical hence
$t=0$ corresponds to a polar angle of $\varphi=0$. If we associate the times of high
harmonic emission with corresponding angles with the lab-frame $x$-axis
we arrive at pairs $\left(t,\varphi\right)=\left(0,0\right)$, $\left(\frac{T}{3},\frac{5\pi}{3}\right)$, $\left(\frac{2T}{3},\frac{4\pi}{3}\right)$, $\left(T,\pi\right)$, $\ldots$.
However, we note that the emission of high-harmonic bursts can occur at both maxima and minima of the driving field.
Since the driving field is linearly polarized in $x$-direction in the rotating frame we can relate these maxima
and minima by considering a mirroring at the rotating-frame $y$-axis. Such a mirroring leads to a shift of the
emission angle in the lab frame by $\pi$. Consequently, taking into account both
emission around a maximum and a minimum of the electric field, we obtain
the following time-angle pairs: $\left(t,\varphi\right)=\left(0,0\right)$, $\left(\frac{T}{3},\frac{2\pi}{3}\right)$, $\left(\frac{2T}{3},\frac{4\pi}{3}\right)$, $\left(T,0\right),\left(\frac{4T}{3},\frac{2\pi}{3}\right)$, $\left(\frac{5T}{3},\frac{4\pi}{3}\right)$, $\left(2T,0\right)$, $\ldots$.
This corresponds precisely to the emission pattern predicted by previous
works~\cite{Milosevic2000,Milosevic2000a} in which in each cycle
of the fundamental driver three linearly polarized emission bursts
are predicted, each rotated by an angle of $120^{\circ}$ with respect
to each other.

\section{Optimizing the HHG Spectrum for Bicircular Driving}\label{sub:bicirc optimise}

For a fixed frequency $\tilde{\omega}$ [Eq.~(\ref{eq:average frequency})]
and field strength $F_{0}$ of the linearly polarized field in the linearizing rotating frame,
any bicircular driving scheme in the lab frame is characterized by a single
parameter: $\alpha$. For $\alpha=0$, we expect due to the linear driving in the rotating frame
a high-harmonic cut-off according to the semiclassical limit at $I_{p}+3.17U_{p}$, with $I_{p}$ the ionization potential and 
$U_{p}=\frac{F_{0}^{2}}{4\tilde{\omega}^{2}}$ the ponderomotive potential. Even
including the effect of the coriolis term, it seems natural to expect that a decrease in $\tilde{\omega}$,
respectively an increase in $F_{0}$, will still lead to an extension of the high-harmonic plateau by increasing
the ponderomotive potential. This leaves the question regarding the role of $\alpha$ on the spectral
cut-off.

\begin{figure}[pt]
\centerline{\includegraphics[width=\columnwidth]{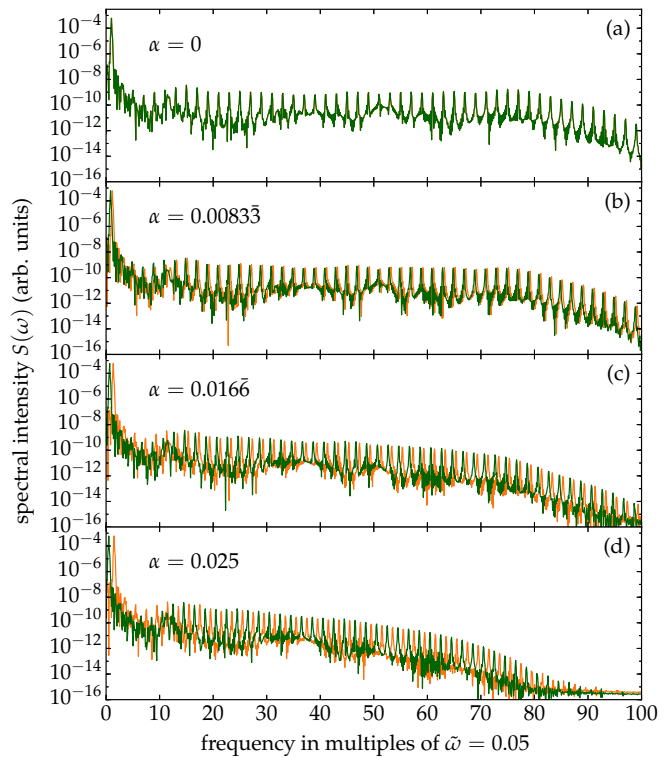}}
\caption{(color online) High-harmonic spectrum in the lab frame for
fixed $\tilde{\omega}$ and $F_0$ and different values of $\alpha$. The green (dark-shaded) curve represents the
right-circular spectrum, while the orange (light-shaded) curve represents the left-circular
spectrum (in panel (a) both are on top since the signal in
$y$-direction vanishes for $\alpha=0$).}
\label{fig:dual_linear_005}
\end{figure}

We show in Fig.~\ref{fig:dual_linear_005}
the high-harmonic spectrum in the lab frame for $\tilde{\omega}=0.05$ and $F_{0}=0.1$
for otherwise identical parameters as in Sec.~\ref{sub:average frequency apply}.
Note that $\tilde{\omega}$ refers to the mean frequency
of the counter-rotating drivers. For $\alpha=0$, cf.~Fig.~\ref{fig:dual_linear_005}(a), lab frame
and rotating frame are identical, and the driving field is linearly
polarized in both. The high-harmonic signal in the $y$-direction
is consequently zero. Hence, the right-circular and left-circular signals are identical
and one clearly observes the expected
behavior of spectral peaks at odd multiples of the laser frequency in the
high-harmonic plateau. The plateau cut-off at around $E_{\text{cutoff}}\simeq80\tilde{\omega}=4.00$
matches perfectly with the semi-classical prediction at $I_{p}+3.17U_{p}=3.97$.

Increasing the value of $\alpha$ gradually depresses the high-harmonic
plateau, cf.~Fig.~\ref{fig:dual_linear_005}(b)-(d). This effect can be
rationalized by a simple semiclassical argument. For higher harmonics
the excursion of the electron needs to be further and further away
from the nucleus, corresponding to a higher energy gain in the electric
field in the framework of the three-step model. However, the larger
the distance from the nucleus the stronger the effect of the $L_{z}$-term
will be which leads to acceleration of the electron in the $y$-direction.
This deflection will reduce the $x$-elongation in favor of the $y$-elongation.
As a consequence the energy acquired by the
electron in the electric field is reduced as it is proportional
to only the $x$-component of its elongation since the laser pulse
is linearly polarized in $x$-direction in the rotating frame. Incidentally,
this also allows for a simple explanation on why counter-rotating
drivers are significantly more suitable than co-rotating drivers if one
aims to obtain circularly polarized high-harmonics. In a co-rotating
scheme $\alpha$ is not given by half of the difference of the frequencies
of the two drivers in the lab frame but by the average of their frequencies.
This immediately implies a much larger value of $\alpha$ leading
to a much stronger depression of the high-harmonic plateau compared
to an equivalent counter-rotating scheme.

Looking at the high-harmonic spectra from Fig.~\ref{fig:dual_linear_005},
shifted according to Eqs.~(\ref{eq:plus signal}) and (\ref{eq:minus signal})
to obtain the circular components of the harmonic signal in the lab
frame, one can observe an interesting dichotomy. On the one hand our
results clearly show that using two drivers that are as close in frequency
as possible leads to the least reduction in the high-harmonic plateau,
on the other hand one requires a sufficiently large $\alpha$ to obtain
spectrally separated signals for the two circular polarization directions
in the lab frame. The width of the harmonic peaks is primarily determined
by the temporal width of the drivers. Hence, the optimal strategy
to obtain spectral selectivity for the two polarization directions
while having a harmonic plateau that is as far extended as possible
consists in choosing $\alpha$ just above the spectral width of the
harmonic peaks. Specifically, for the typically chosen `fundamental plus second harmonic' driving
scheme, shown in Fig.~\ref{fig:dual_linear_005}(c), we observe
that our particular chosen set of parameters would still allow closer frequencies for the
bicircular drivers while maintaining a sufficient
spectral separation of the right-circular and left-circular harmonic
peaks, cf.~Fig.~\ref{fig:dual_linear_005}(b).

From an experimental point of view this means that for driving pulses
with several optical cycles the best results will be obtained by choosing
driving fields originating from a laser with frequency $\omega_{0}$
where the left-circularly polarized driver is tuned lower by a small
amount while the right-circularly polarized driver is tuned higher
by the same amount (or vice versa). Conversely, if the driving fields
consist of only very few cycles then the spectral width of the peaks
will increase and it is likely that choosing low-order harmonics of
a fundamental frequencies is a more suitable approach. In particular,
employing a left-circularly polarized driver at some fundamental
$\omega_0$ and a right-circularly polarized driver at the third harmonic
$3\omega_0$ leads to $\tilde{\omega}=2\omega_0$ and a maximal spectral separation
of the peaks in the high-harmonic spectrum given by $\tilde{\omega}$.
However, this comes with a cost, namely a coriolis term in the rotating
frame with strength $\alpha=\omega_0$.
This represents already a moderately strong depression of the high
harmonic plateau, cf. Fig.~\ref{fig:dual_linear_005}(d).

\section{Perspectives for Rotating Frame Analyses I: Elliptical drivers \label{sec:perspectives I}}

We can generalize Eq.~(\ref{eq:bicirc_start}) to the case of one
of the two drivers having elliptical polarization instead of circular
polarization, a situation that has been recently explored experimentally \cite{Fleischer2014}
and also discussed from a theoretical point of view~\cite{Pisanty2014,Milosevic2015}. 
The Hamiltonian in this case can be written as
\begin{eqnarray}
\nonumber H\left(t\right) & = & H_{0}+F_{0}\left(t\right)\left(A\left(\epsilon\right)x\cos\left(\omega_{1}t\right)+B\left(\epsilon\right)y\sin\left(\omega_{1}t\right)\right. \\
                          &   & \left.+x\cos\left(\omega_{2}t\right)-y\sin\left(\omega_{2}t\right)\right)\,,
\end{eqnarray}
where $A\left(\epsilon\right)=\frac{\sqrt{2}}{\sqrt{1+\epsilon^{2}}}$
and $B\left(\epsilon\right)=\frac{\sqrt{2}\epsilon}{\sqrt{1+\epsilon^{2}}}$,
with $\epsilon$ being the ellipticity of the field with frequency
$\omega_{1}$. 
The case $\epsilon=1$ leads back to the case of bicircular driving with equal strengths
whereas $\epsilon=0$ corresponds to the superposition of a linearly polarized field with
a left-circularly polarized one.
Following the steps from Sec.~\ref{sec:linearising rotating frame},
by going into a rotating frame with frequency $-\alpha$ we arrive
at the intermediate expression
\begin{eqnarray*}
H' & = & H_{0}+\alpha L_{z} \\
   &   & \hspace{-8ex}+F_{0}\left(t\right)\left\{ \frac{1}{2}\left(A+B\right)\left(x\cos\left(\left[\alpha+\omega_{1}\right]t\right) + y\sin\left(\left[\alpha+\omega_{1}\right]t\right)\right) \right. \\
   &   & \hspace{-8ex}+\frac{1}{2}\left(A-B\right)\left(x\cos\left(\left[\omega_{1}-\alpha\right]t\right)-y\sin\left(\left[\omega_{1}-\alpha\right]t\right)\right) \\
   &   & \hspace{-8ex}\left.\vphantom{\frac{1}{2}}+x\cos\left(\left[\omega_{2}-\alpha\right]t\right)-y\sin\left(\left[\omega_{2}-\alpha\right]t\right)\right\} \,.
\end{eqnarray*}
For brevity we suppress the $\epsilon$ dependence of $A$ and $B$.
An `on-frequency' co-rotating frame is obtained by choosing $\omega_{1}=\alpha$
leading to the rotating-frame Hamiltonian
\begin{eqnarray}
\nonumber H' & = & H_{0}+\omega_{1}L_{z}+\frac{1}{2}\left(A-B\right)F_{0}\left(t\right)x \\
\nonumber    &   & +F_{0}\left(t\right)\left\{ \frac{1}{2}\left(A+B\right)\left(x\cos\left(2\omega_{1}t\right)+y\sin\left(2\omega_{1}t\right)\right) \right.\\
             &   & \left.\vphantom{\frac{1}{2}}+x\cos\left(\left[\omega_{2}-\omega_{1}\right]t\right)-y\sin\left(\left[\omega_{2}-\omega_{1}\right]t\right)\right\} \,, \label{eq:h_ellipcirc}
\end{eqnarray}
which can be interpreted as the system being subject to two counter-rotating
circular drivers with different amplitude
in the presence of a coriolis term and a static electric field. While this
field is not truly static since it is modified by the pulse envelope the timescale
of this modulation is slow enough such that the electric field can be regarded as static
for the purposes of HHG except for extremely short drivers.
For $A=B$ ($\epsilon=1$) we reobtain the case
from Sec.~\ref{sec:linearising rotating frame} which manifests
through a vanishing static electric field term in the rotating frame. 

The shift property for the HHG spectra between rotating frame and lab frame
of circular emission components, cf.~Sec.~\ref{sec:hhg_rot_transform}, allows once 
again to use symmetry arguments in the rotating frame to obtain information about the actual
HHG spectrum in the lab frame. For $A=B$ it can easily be seen from Eq.~(\ref{eq:h_ellipcirc})
that in the time-independent Hamiltonian $L_{z}$ is conserved which means that emission can only take place
with $\left(n+1\right)$ photons from the right-circular driver and
$n$ from the left-circular driver, respectively $\left(n-1\right)$
photons from the right-circular driver and $n$ photons from the left
circular driver \cite{Pisanty2014,Kfir2015,Milosevic2015}. This leads in the first case to right-circularly
polarized emission at frequencies $n\left(\omega_{1}+\omega_{2}\right)+\left(\omega_{2}-\omega_{1}\right)$,
corresponding to a right-circularly polarized field in the lab frame
at $n\left(\omega_{1}+\omega_{2}\right)+\omega_{2}-2\omega_{1}$,
respectively left-circularly polarized emission at frequencies $n\left(\omega_{1}+\omega_{2}\right)-\left(\omega_{2}-\omega_{1}\right)$,
corresponding to a left-circularly polarized field in the lab frame
at $n\left(\omega_{1}+\omega_{2}\right)-\left(\omega_{2}-2\omega_{1}\right)$.

Once $A$ starts to deviate from $B$ both a 
breaking of inversion symmetry and a breaking of
conservation of the angular momentum occur in the `static' Hamiltonian.
This opens up arbitrary combinations of the driving fields at frequencies $\left(2n-m\right)\omega_{1}+m\omega_{2}$
where $n>m$ corresponds to right-circular signals, which appear in
the lab frame at $\left(2n-m-1\right)\omega_{1}+m\omega_{2}$, and
$n<m$ corresponds to left-circular signals, which appear in the lab
frame at $\left(2n-m+1\right)\omega_{1}+m\omega_{2}$. The case $n=m$
corresponds to linear emission channels at frequencies $n\left(\omega_{1}+\omega_{2}\right)$ 
with equal contributions from the counter-rotating circular drivers in the rotating frame.
This channel will split in the lab frame into two signals:
a right-circularly polarized signal at $n\left(\omega_{1}+\omega_{2}\right)-\omega_{1}$
and a left-circularly polarized signal at $n\left(\omega_{1}+\omega_{2}\right)+\omega_{1}$.
Note that the larger the discrepancy
between $A$ and $B$ the more dominant contributions with greater difference between $n$ and
$m$ will become. This is because the intensity ratio between the $n$-right-circular
driver and $m$-left-circular driver in the rotating frame goes as 
$\frac{I_{n}}{I_{m}}=\frac{\left(A+B\right)^{2}}{4}=\frac{1}{2}+\frac{\epsilon}{1+\epsilon^2}$
which is a monotonously increasing function for $\epsilon\in\left[-1,1\right]$.
Additionally, the more $A$ and $B$ differ the
stronger the static field term becomes which leads to larger and larger
symmetry breaking, washing out the HHG among more and more channels.

In light of the importance of the difference between $n$ and $m$
it is useful to define $\bar{n}\equiv n-m$. Then, we can state the
following: One observes left circularly polarized signals in the lab
frame at frequencies $\left(2\bar{n}+m-1\right)\omega_{1}+m\omega_{2}$
for $\bar{n}\geq0$ and right-circularly polarized signals in the
lab frame at frequencies $\left(2\bar{n}+m+1\right)\omega_{1}+m\omega_{2}$
for $\bar{n}\leq0$. The principal order of the signal is given by
$m$ whereas $\left|\bar{n}\right|$ reflects the required symmetry
breaking where the breaking strength is proportional to the deviation of
$\left|\bar{n}\right|$ from $1$, with only $\left|\bar{n}\right|=1$ corresponding 
to symmetry conservation. These findings match perfectly
with those in Ref.~\cite{Milosevic2015}.

\section{Perspectives for Rotating Frame Analyses II: Static Electric and Magnetic Fields \label{sec:perspectives II}}

In addition to the appearance of a static electric field term by entering a rotating frame
as in Sec.~\ref{sec:perspectives I} there is also an interesting connection
to static magnetic fields for both the rotating-frame descriptions obtained in
Secs.~\ref{sec:linearising rotating frame} and \ref{sec:perspectives I}.
This becomes clear when considering an axially symmetric system under
the influence of a static magnetic field along the $z$-direction,
$\vec{B}=B_{0}\vec{e}_{z}$. The full Hamiltonian in Coulomb gauge
reads in this case
\begin{eqnarray}
H_{\text{mag}} & = & H_{0}+\frac{B_{0}}{2}L_{z} + \frac{B_{0}^{2}}{8}\left(x^{2}+y^{2}\right)\,.\label{eq:statmag}
\end{eqnarray}
The angular momentum term appearing in this context has an identical form to the coriolis term
obtained for rotating-frame Hamiltonians, but there is an additional contribution
that leads to harmonic trapping along the $z$-axis proportional in strength
to the square of the magnetic field. It should be pointed out, however,
that this term does preserve the axial symmetry of $H_{0}$
and thus will not alter the harmonic spectrum in terms of symmetry-forbidden harmonics.
Still, similarly to the coriolis term, the signal strength of the emitted higher harmonics
will be reduced by this term since long excursions from the nucleus are suppressed.
Furthermore a `true' static magnetic field in the lab frame readily allows to tune
the harmonic term and the coriolis term independently since the strength of
the harmonic term is invariant under a transformation to the rotating frame while the
total coriolis term will have strength $B_{0}+\alpha$ where $\alpha$ is the rotation
frequency of the rotating frame. This allows to independently study the effects of both
contributions in a practical setup.

We want to briefly address the fact that the initial state for HHG can be
different in the magnetic field case compared to the case of merely
considering a rotating frame. This can for example occur when the
presence of the magnetic field breaks the symmetry in a degenerate ground state
manifold. In this case the ground state of $H_{\text{mag}}$
will not follow the symmetries of $H_{0}$. This may
significantly alter the HHG spectrum as high-harmonic spectra 
show a clear dependence on the initial state of the system, see, e.g., Ref.~\cite{Medisauskas2015} for
an analysis with respect to bicircular drivers. For high magnetic fields it is also conceivable
that an excited state with, e.g., $p$ character becomes energetically
lowered below the energy of the ground state with $s$ character (which
is unaffected by a Zeeman splitting). However, if the magnetic field
is established sufficiently slowly the ground state will likely remain
invariant due to adiabatic following. For the case of static electric
fields very similar arguments hold.

Previous theoretical studies in terms of static electric and magnetic fields concluded that
an appreciable impact on the HHG spectra can only be observed for field strengths at the
very edge of experimental feasibility \cite{Milosevic1999,Bandrauk2003}. In a rotating-frame picture these field strengths are
achieved rather naturally since they originate from AC field strengths (for electric fields),
respectively from the frequency of the rotating-frame description (for magnetic fields). For example,
using an $800$ nm right-circularly polarized driver and its second harmonic at $400$~nm as a left-circularly
polarized driver, the linearizing rotating frame as discussed in Sec.~\ref{sec:linearising rotating frame} leads to
a coriolis term which would correspond by Eq.~(\ref{eq:statmag}) to a static magnetic field of almost $12000$~T.

Circular high-harmonic emission in the lab frame and the rotating frame
are directly related to each other by only a shift in frequency, cf.~Sec.~\ref{sec:hhg_rot_transform}.
This allows on the one hand to directly apply knowledge about static-field effects to understanding high
harmonic spectra of circular light while on the other hand enabling the possibility to engineer Hamiltonians
in the rotating frame that realize the effects of such static fields even at field strengths beyond
current technical feasibility.

\section{Conclusions \label{sec:conclusions}}

HHG of circularly polarized light with bicircular drivers can be understood in
many ways, ranging from a purely photonic picture to a strictly classical four-wave mixing interpretation
\cite{Pisanty2014}. A particularly simple way to approach this process is by moving to a rotating frame
of reference where the bicircular driving reduces to a linearly polarized field and a coriolis term
\cite{Bandrauk2003,Yuan2015}. We showed in this paper that the circularly polarized components of the
high-harmonic signal in the lab frame follow directly from their counterparts in the rotating
frame under a simple shift in frequency. This allows to directly deduce properties of those spectra and
even the generation process itself from the well-studied HHG with linearly polarized driving fields. The
influence of the coriolis term for axially symmetric systems on the high-harmonic plateau consists
in a reduction of the cutoff frequency proportional to the strength of this term. Otherwise
it does not alter the peak structure in the spectrum due to the preservation of the axial symmetry.
As a consequence it is possible to determine the optimal bicircular drivers when the goal is to preserve
a high-harmonic plateau that is extended as far as possible - the drivers need to be counter-rotating and
their frequencies should be chosen as similarly as possible. Only the need for a spectral separation
of the two circularly polarized components requires to choose a nonzero difference in frequency to
compensate for the finite width of the peaks in frequency domain due to the finite width of the
driving fields in time. Conversely, if such a spectral separation is not necessary because the right and
left-circularly polarized signal can be otherwise filtered out, choosing identical frequencies is clearly
the best approach. This is an observation that very recently has been made in an experimental study where
circularly polarized high harmonics have been generated in a non-collinear scheme allowing for a spatial
separation of the two circularization directions \cite{Hickstein2015}.

Beyond the case of bicircular drivers, a rotating-frame picture also offers a valuable perspective on elliptical
drivers. The occurrence of additional peaks in the high-harmonic spectrum when moving away from the bicircular
case originates in a properly chosen rotating frame from the presence of an additional static electric field. This static electric field breaks
the symmetries present for bicircular drivers and opens up additional channels for HHG
in the rotating frame that can then be directly observed in the lab frame taking into account the corresponding
shift in frequency. Furthermore, the rotating-frame description of these schemes shows a near-duality to
HHG under static electric and magnetic fields. While the effects of such fields have been
examined in the past, the regime in which a visible impact on the spectra could be experimentally observed
is mostly inaccessible due to the infeasibility of generating the corresponding strong fields. However, the
rotating-frame picture clearly illustrates that the effect of such strong fields appear naturally in a
suitably chosen frame of reference and the resulting modifications of the high-harmonic spectra of these
`virtual' fields have clearly observable consequences in experiments that can readily be performed today.

The rotating-frame picture proves to be a powerful tool in understanding HHG of
circularly polarized light. Depending on the particular choice of driving fields it enables a close correspondence to
linearly polarized drivers and even HHG in the presence of static electric and magnetic
fields. It can thus serve as a pivotal link between seemingly disjoint setups for HHG
and allows for a remarkably simple explanation of many properties of the experimentally observed spectra.

\begin{acknowledgments}
  This work was supported by the European Research Council StG (Project No.~277767-TDMET) 
  and the VKR center of excellence, QUSCOPE.
  D.M.R.~gratefully acknowledges support from the
  Alexander-von-Humboldt foundation through the Feodor Lynen program.
\end{acknowledgments}

\appendix

\section{HHG Spectra via Dipole Acceleration in Rotating Frames \label{sec:Dipole acceleration}}

\begin{figure}[pt]
\centerline{\includegraphics[width=\columnwidth]{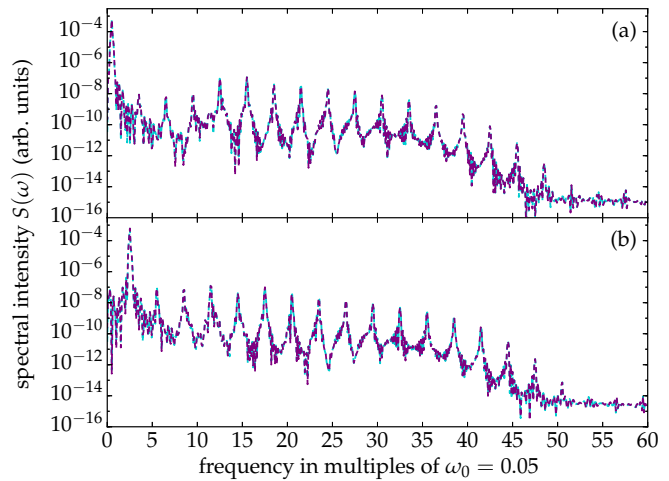}}
\caption{(color online) (a) High-harmonic spectrum for the right-circularly polarized
field component and (b) the left-circularly polarized field component
for two counter-rotating fields with frequency $\omega_{1}=\omega_{0}=0.05$,
respectively $\omega_{2}=2\omega_{0}=0.1$, with equal field
amplitude of $F_{0}=0.05$. The driving laser pulse is trapezoidally
shaped with the same parameters as in Fig.~\ref{fig:peakana}. The
turquoise (light-shaded) curves are the results in the lab frame, the purple (dark-shaded)
curves are the results in a rotating frame with rotation frequency $\alpha=0.025$
shifted to the left by $\alpha$ for right-circular polarization,
respectively to the right by $\alpha$ for left-circular polarization.}
\label{fig:numerical_acceleration}
\end{figure}

The expectation value of the dipole acceleration can be computed via
Ehrenfest's theorem,
\[
\left\langle \vec{a}\right\rangle =\frac{d^{2}}{dt^{2}}\left\langle \vec{x}\right\rangle =\frac{d}{dt}\left\langle \vec{p}\right\rangle =-\left\langle \vec{\nabla} V\right\rangle \,.
\]
We can then perform the following calculation
\footnote{Note that we neglect the contribution from the driver since it
will not have any impact on the higher harmonics. Furthermore, we
neglect the minus sign from the electronic charge since it will not
impact the actual harmonic signals.},
\begin{eqnarray*}
\left\langle \vec{a}\right\rangle _{\text{lab}} & = & \frac{d}{dt}\left\langle \vec{p}\right\rangle _{\text{lab}} \\
                                                & = & \frac{d}{dt}\left\langle U^{\dagger}\left(t\right)U\left(t\right)\vec{p}U^{\dagger}\left(t\right)U\left(t\right)\right\rangle _{\text{lab}}\\
                                                & = & \frac{d}{dt}\Braket{\psi_{\text{rot}}\left(t\right)|e^{-i\alpha tL_{z}}\vec{p}e^{i\alpha tL_{z}}|\psi_{\text{rot}}\left(t\right)}\\
                                                & = & i\Braket{\psi_{\text{rot}}\left(t\right)|\left[H_{\text{rot}},e^{-i\alpha tL_{z}}\vec{p}e^{i\alpha tL_{z}}\right]|\psi_{\text{rot}}\left(t\right)} \\
                                                &   & +\Braket{\psi_{\text{rot}}\left(t\right)|e^{-i\alpha tL_{z}}\left[-i\alpha L_{z},\vec{p}\right]e^{i\alpha tL_{z}}|\psi_{\text{rot}}\left(t\right)}\\
                                                & = & i\Braket{\psi_{\text{rot}}\left(t\right)|\left[H_{0}+\alpha L_{z},e^{-i\alpha tL_{z}}\vec{p}e^{i\alpha tL_{z}}\right]|\psi_{\text{rot}}\left(t\right)}\\
                                                &   & -i\alpha\Braket{\psi_{\text{rot}}\left(t\right)|e^{-i\alpha tL_{z}}\left[L_{z},\vec{p}\right]e^{i\alpha tL_{z}}|\psi_{\text{rot}}\left(t\right)}\\
                                                & = & i\Braket{\psi_{\text{rot}}\left(t\right)|\left[H_{0},e^{-i\alpha tL_{z}}\vec{p}e^{i\alpha tL_{z}}\right]|\psi_{\text{rot}}\left(t\right)}\,.
\end{eqnarray*}
For the right-circular component of the acceleration this means that
\begin{eqnarray*}
\left\langle a_{+}\right\rangle _{\text{lab}}  &   & \\
 &   & \hspace{-8ex} =i\frac{1}{\sqrt{2}}\Braket{\psi_{\text{rot}}\left(t\right)|\left[H_{0},e^{-i\alpha tL_{z}}\left(p_{x}+ip_{y}\right)e^{i\alpha tL_{z}}\right]|\psi_{\text{rot}}\left(t\right)}\\
 &   & \hspace{-8ex} =i\frac{e^{-i\alpha t}}{\sqrt{2}}\Braket{\psi_{\text{rot}}\left(t\right)|\left[H_{0},p_{x}+ip_{y}\right]|\psi_{\text{rot}}\left(t\right)}\\
 &   & \hspace{-8ex} =-\frac{1}{\sqrt{2}}e^{-i\alpha t}\left[\left\langle \nabla_{x}V\right\rangle _{\text{rot}}+i\left\langle \nabla_{y}V\right\rangle _{\text{rot}}\right]\\
 &   & \hspace{-8ex} =-e^{-i\alpha t}\left\langle \nabla_{+}V\right\rangle _{\text{rot}}\,,
\end{eqnarray*}
where $V$ is the potential term in the Hamiltonian $H_{0}$ and $\nabla_{\pm}\equiv\frac{1}{\sqrt{2}}\left(\nabla_{x}\pm i\nabla_{y}\right)$.

As a consequence one obtains the high-harmonic signal for the right-circular
component via
\begin{eqnarray*}
S_{+}^{\text{lab}}\left(\omega\right) & = & \left|\int\left\langle a_{+}\right\rangle _{\text{lab}}e^{-i\omega t}\ dt\right|^{2}\\
                                      & = & \left|\int\left\langle \nabla_{+}V\right\rangle _{\text{rot}}e^{-i\left(\omega+\alpha\right)t}\ dt\right|\\
                                      & = & S_{+}^{\text{rot}}\left(\omega+\alpha\right)\,,
\end{eqnarray*}
which represents a shift of the high-harmonic signal in right-circular
direction obtained in the rotating frame in direct analogy to the case for
the expectation value of the dipole, cf.~Eq.~(\ref{eq:plus signal}).
The derivation for the left-circular direction can be performed analogously 
leading to the respective result as in Eq.~(\ref{eq:minus signal}).
An important point
to emphasize is that when using Ehrenfest's theorem in the rotating
frame one needs to compute the derivative of the potential term corresponding to
the non-rotating Hamiltonian, i.e., the coriolis term does not enter.

A comparison of the high-harmonic spectrum obtained via Ehrenfest's
theorem in the rotating frame versus the lab frame is shown in
Fig.~\ref{fig:numerical_acceleration}. Evidently, both results match
extraordinarily well. The remaining discrepancies stem from the fact
that the spectral grid points of the shifted spectrum from the rotating
frame and those of the unshifted spectrum from the lab frame do not quite match
since the time grid we employed leads to a distance between two grid points 
in frequency domain that is no integer multiple of $\alpha$.

\end{document}